\documentstyle[aps,multicol,psfig]{revtex}
 \draft
\tightenlines
\begin{document}
\title{Signature of Chaotic Diffusion in Band Spectra}
\author{T.~Dittrich, B.~Mehlig, H.~Schanz}
\address{Max-Planck-Institut f\"ur Physik komplexer Systeme,
N\"othnitzer Stra\ss e 38, D-01\,187 Dresden, Germany}
\author{U.~Smilansky}
\address{Department of Physics of Complex Systems,
Weizmann Institute of Science, Rehovot 76\,100, Israel}
\date{September 29, 1997}
\maketitle

\begin{abstract}
We investigate the two-point correlations in the band spectra of periodic
systems that exhibit chaotic diffusion in the classical limit, in terms of
form factors with the winding number as a spatial argument. For times below
the Heisenberg time, they contain the full space-time dependence of the
classical propagator. They approach constant asymptotes via a regime,
reflecting quantal ballistic motion, where they decay by a factor proportional
to the number of unit cells. We derive a universal scaling function for the
long-time behaviour. In the limit of long chains, our results are consistent
with expressions obtained by field-theoretical methods. They are substantiated
by numerical studies of the kicked rotor and a billiard chain.
\end{abstract}
\begin{multicols}{2}
\section{Introduction}
A large class of systems, among them most prominently solid-state systems, are
organized as repetitions of identical or near-identical units. If in such a
system the {\it classical\/} dynamics is chaotic and the unit cells are
connected, then globally this leads to a diffusive spreading of trajectories,
irrespective of the presence or absence of static disorder. Regular motion on
tori does not contribute to diffusive spreading: Closed tori do not allow for
transport along the lattice, while open tori give rise to ballistic spreading.
The spectral and transport properties of extended {\it quantum\/} systems, in
contrast, depend sensitively on the degree of translational symmetry. On short
time scales, however, where the quantum dynamics still closely follows the
classical, both periodic and disordered quantum systems exhibit (apart from an
initial ballistic phase which is of no interest for the following) a spreading
of wavepackets with the characteristics of the chaotic diffusion in their
classical counterparts. The signature of this phase in the discrete spectra of
disordered systems with Anderson-localized eigenstates has been investigated
previously \cite{dit1,arg}, using techniques in the spirit of Berry's
semiclassical derivation \cite{han,ber} of the spectral form factors for the
canonical random-matrix ensembles.

In the present work, we report on a study of periodic systems with band
spectra and eigenstates of Bloch form, focusing on the signature of spatial
order and dynamical disorder in the spectral two-point correlations. We
exploit the existence of a second conserved quantum number besides the energy,
the quasimomentum, to define form factors in the canonically conjugate space
spanned by time and winding number. They are related to the spatially
coarse-grained propagator and therefore ideally suited to extract dynamical
information from the band structure, without recurring to a local
spectrum. Here, we evaluate this relation over the entire time evolution of
the form factors. In the semiclassical regime, we show that the full space and
time dependence of the classical propagator is contained in the form
factors. Chaotic diffusion in spatially periodic quantum systems can thus
directly be identified in the band structure. Similarly, the spectral
signature of the crossover to quantum ballistic motion can be analyzed on
basis of the full quantum propagator. In the limit of a large number of unit
cells, the form factors exhibit a marked peak in the vicinity of the
Heisenberg time, a consequence of the clustering of levels in quasicontinuous
bands. We obtain a universal scaling function for the long-time behaviour.

Form factors containing the correlations of levels across the Brillouin zone
have previously been studied using the supermatrix nonlinear sigma model
\cite{sim,tan}. Our approach is complementary in that it emphasizes the
important concept of winding numbers, providing direct access to spatial
information.  In combination with semiclassical techniques, this allows to
draw a particularly transparent picture of the physics over all time
regimes. We shall demonstrate that in the case of diffusive spreading and in
the limit of an infinite number of unit cells, our theory allows to reproduce
the principal results of the sigma-model approach. At the same time, however,
we are able to go beyond those results in that we need not make any
assumptions as to the number of unit cells in the system, nor on the mode of
density relaxation.

Our theory therefore applies to a broad class of systems. They include
semiconductor superlattices supporting chaotic electron motion (``antidot
arrays'') \cite{gei,muc,sil,lun}, quantum-optical systems involving
periodically modulated standing-wave fields \cite{kol}, and KAM systems with
toroidal chaotic layers in phase space containing chains of regular resonance
islands, as they occur frequently in molecules \cite{kor}. For this type of
systems, there exists a large body of spectral data, both experimental and
numerical \cite{gei,muc,sil,lun,kol,kor}. By analyzing these data as explained
in the following, the dynamical information encoded in the respective band
structures can be extracted.

\section{The Generalized Form Factor}
The setup we have in mind is a finite chain of $N$ identical unit cells, with
cyclic boundary conditions at its ends. Spatial disorder within the unit cell
is {\it not\/} required. We restrict ourselves to quasi-one-dimensional
lattices, since the generalization to higher dimensions is straightforward.

As a consequence of periodicity, the spectrum can be decomposed into $N$
subsets, each of which corresponds to one of the $N$ irreducible
representations of the group of lattice translations $\widehat T(na{\bf
e}_x)$, $n$ integer. Together, these subspectra coalesce into discretized
bands and approach continuous bands in the limit $N \rightarrow \infty$.
Dynamical and spectral quantities specific to one of the irreducible
representations \cite{rob,cre} are constructed using the projectors
\begin{equation}
\widehat P_m = \frac{1}{N} \sum_{n=0}^{N-1}
\chi_m(n) \widehat T^{\dagger}(na{\bf e}_x)
\end{equation}
onto the corresponding subspaces. They invoke the group characters $\chi_m(n)
= \exp({\rm i}n\theta_m)$. We refer to the $\theta_m = 2\pi m/N$, $m = 0,\,
\ldots,\, N-1$, as Bloch phases. The symmetry-projected Green function is
defined as
\begin{equation}
\widehat G_m(E) = \widehat P_m \widehat G(E),
\label{blochgreen}
\end{equation}
where $\widehat G(E)$ is the Green function for the full chain. From $\widehat
G_m(E)$, other Bloch-phase-specific quantities can be derived as if they
pertained to the full spectrum of a system without spatial symmetry. For
example, the Bloch-phase-specific spectral density is related to the trace of
the corresponding Green function, Eq.~(\ref{blochgreen}), in the usual way
(see, e.g., Ref.~\cite{gut}),
\begin{equation}
d_m(E) = \sum_\alpha \delta(E\!-\! E_\alpha(\theta_m))
= -\frac{1}{\pi}{\rm Im}\,{\rm tr}\,[\widehat G_m(E)].
\end{equation} 
Here, the trace extends only over a single unit cell.

The basic energy scale in the following is the inverse mean spectral density
{\em per unit cell}, $1/\langle d_{\rm uc} \rangle$, or equivalently, the mean
separation of neighbouring bands. We define also the Heisenberg time with
respect to the unit cell, $t_{\rm H} = 2\pi\hbar\langle d_{\rm uc}
\rangle$. Accordingly, we scale time as $\tau = t/t_{\rm H}$ and energy as $r
= \langle d_{\rm uc}\rangle E$. In these units, the size of the spectral
window considered is $\Delta r$, roughly the total number of bands.

The time-domain counterpart of $d_m(E)$ is the amplitude 
\begin{equation}
a_m(\tau) =
\int_{-\infty}^{\infty} {\rm d}r\, {\rm e}^{-2\pi{\rm i} r \tau}
d_m(r/\langle d_{\rm uc}\rangle).
\end{equation} 
By performing another, now discrete, Fourier transform with respect to
$\theta_m$ \cite{cre,scha}, which amounts to going from the Bloch-phase to the
winding-number representation, we define the amplitude
\begin{eqnarray}
\tilde a_n(\tau) &=& \frac{1}{N} \sum_{m=0}^{N-1}
\exp({\rm i}n\theta_m) a_m(\tau) \nonumber\\
&=& \int_{\rm unit\;cell} {\rm d}q\,
\langle{\bf q} + na{\bf e}_x|\widehat U(\tau t_{\rm H})|{\bf q}\rangle,
\label{eq1}
\end{eqnarray}
where $a{\bf e}_x$ generates the lattice, $|{\bf e}_x| = 1$.
Winding-number-specific form factors are defined as
\begin{equation}
\widetilde K_n(\tau) = \frac{1}{\Delta r} |\tilde a_n(\tau)|^2\,.
\end{equation}
Substituting Eq.~(\ref{eq1}) shows that
the $\widetilde K_n(\tau)$ comprise pairs of levels with all Bloch phases. In
particular, $\widetilde K_0(\tau)$ corresponds to the form factor for the
entire spectrum, irrespective of spatial periodicity.  Eq.~(\ref{eq1})
represents a partial trace of the propagator $\widehat U$. The $\widetilde
K_n(\tau)$ can therefore be interpreted as probabilities to return after
encircling the unit cell $n$ times.

\section{Semiclassical Regime}
A semiclassical account of the symmetry-projected spectral quantities is
achieved on the basis of a generalized concept of periodic orbits
\cite{rob,leb,scha}. It becomes transparent in a symmetry-reduced
representation of the chain, where the two boundaries of the unit cell
connected by the translational symmetry are identified. In this way, the unit
cell assumes the topology of an annulus, possibly times additional dimensions.
An orbit periodic in this reduced space can be classified according to its
topology, expressed by its winding number, the number of times it runs around
the unit cell before closing. The contribution of a periodic orbit $j$ to
$d_m(E)$ contains the additional phase $-n_j\theta_m$, where $n_j$ is the
winding number of this orbit \cite{rob,leb,scha}. By the two Fourier
transforms that lead from $d_m(E)$ to $\tilde a_n(\tau)$, we obtain the
semiclassical trace formula
\begin{eqnarray}
\tilde a_n(\tau) = \sum_j &&
\frac{\tau_j^{({\rm p})}}{\sqrt{|\det({\sf M}_j - {\sf I})|}}
\exp\left({\rm i}\frac{S_j(E)}{\hbar} + {\rm i}\mu_j\frac{\pi}{2}\right)
\nonumber\\
&& \times\,\delta_{1/\Delta r}(\tau - \tau_j) \delta_{(n-n_j)\,{\rm mod}\, N},
\end{eqnarray}
with $\tau_j^{({\rm p})}$, ${\sf M}_j$, $S_j$, and $\mu_j$ denoting,
respectively, primitive period, monodromy matrix, action, and Maslov index of
the periodic orbit $j$. Besides the usual amplitude and phase factors, this
trace formula has attained two delta functions: A broadened $\delta_{1/\Delta
r}(\tau - \tau_j)$ of width $1/\Delta r$ picks out orbits with scaled period
$\tau_j \approx \tau$. An $N$-periodic Kronecker delta $\delta_{(n-n_j)\,{\rm
mod}\, N}$ selects periodic orbits with a winding number that differs from $n$
at most by an integer multiple of $N$.

According to the above interpretation, the $\widetilde K_n(\tau)$, for $\tau <
1$, should be related to the classical probabilities $P_n^{({\rm cl})}(t)$ to
return, after $n$ windings, in the symmetry-reduced phase space. Indeed,
within the diagonal approximation \cite{dit1,arg,dit2}, we derive
\begin{equation}
\widetilde K_n^{({\rm sc})}(\tau) =
\gamma_n \tau P_n^{({\rm cl})}(\tau t_{\rm H}),
\qquad \mbox{$\tau < 1$.}
\label{eq2}
\end{equation}
Here, we have neglected the contribution of repetitions of shorter periodic
orbits. We have not taken the occurrence of self-retracing orbits into account
in order to replace individual degeneracy factors, expressing time-reversal
(${\sf T}$) invariance, by a global $\gamma_n$. Reflecting weak localization
as a function of $n$, it takes the value 2 if orbits with $n_j = n$ are
generically ${\sf T}$ degenerate and 1 otherwise. The winding-number
representation thus enables a direct and natural access to weak-localization
enhancements in the form factor. In the Bloch-phase representation, by
contrast, weak localization is reflected in a smooth transition from GOE
statistics near the symmetry points of the Brillouin zone to GUE statistics
elsewhere \cite{sim,muc,sil,dit2}.

In order that an orbit contribute to $P_n^{({\rm cl})}(t)$, it must be
periodic up to a lattice translation by $na{\bf e}_x$. Assuming that the long
periodic orbits spread as the generic, non-periodic ones, we express
$P_n^{({\rm cl})}(t)$ in terms of the full classical propagator $p({\bf
r}',{\bf r};t)$ as a partial trace,
\begin{equation}
P_n^{({\rm cl})}(t) = \int_{\rm unit\;cell} {\rm d}r\,
p\big({\bf r} + na({\bf 0},{\bf e}_x),{\bf r};t\big),
\label{eq3}
\end{equation}
where ${\bf r} = ({\bf p},{\bf q})$ denotes a point within the unit cell on
the energy shell \cite{han}. Equations (\ref{eq2}), (\ref{eq3}) show that the
generalized form factors, in the semiclassical regime, relate the band
structure to the full, if coarse grained, classical propagator.

The validity of Eq.~(\ref{eq2}) is not restricted to any specific form of
relaxation of the classical distribution, provided the underlying classical
dynamics is predominantly chaotic. For example, billiard chains connected only
by narrow bottlenecks show a marked deviation from normal diffusion on the
time scale of the escape from a single cell. The generalization to
higher-dimensional lattices is straightforward. Also there, $P_n^{({\rm
cl})}(t)$ takes forms significantly different from diffusion in one dimension.

As a specific example, we evaluate Eq.~(\ref{eq2}) for normal diffusion in one
extended dimension. For an interval of length $L = Na$ with cyclic boundary
conditions, the diffusion equation is solved by the propagator $p(x',x;t) =
{\cal G}^{({\rm mod}\, L)}(x'-x,Dt)$, where ${\cal {G}}^{({\rm mod}\,
p)}(x,\sigma^2)$ denotes a normalized Gaussian of period $p$ and variance
$\sigma^2$. We assume that in the nonperiodic dimensions of phase space, the
relaxation towards equidistribution is rapid on the relevant time scales. For
times $t \ll t_{\rm d} = L^2/\pi D$, the Thouless time for the full chain of
length $L$, diffusion is free, $p(x',x;t) = (2\pi Dt)^{-1/2}
\exp(-(x'-x)^2/2Dt)$, while for $t \gg t_{\rm d}$, equidistribution $p(x',x;t)
= 1/L$ is approached. If, as it is the case here, $p(x',x;t) = p(x'-x;t)$, the
partial trace of the propagator amounts to multiplication by the cell size. We
find for $\tau < 1$,
\begin{eqnarray}
\widetilde K_n^{({\rm sc})}(\tau) &=& \frac{\gamma_n\tau}{N}\,
{\cal G}^{({\rm mod}\, 1)}\left(\frac{n}{N},
\frac{g_{\rm uc}\tau}{\pi N^2}\right)\nonumber\\
&=& \left\{ \begin{array}{l@{\quad}l}
\gamma_n\sqrt{\tau/2g_{\rm uc}} {\rm e}^{-\pi n^2 /2g_{\rm uc}\tau}
& \tau \ll N^2/g_{\rm uc},\\
\gamma_n\tau/N & \tau \gg N^2/g_{\rm uc},
\end{array} \right.
\label{eq4}
\end{eqnarray}
introducing the dimensionless parameter $g_{\rm uc} = N^2\times$ $t_{\rm
H}/t_{\rm d} = 2\pi^2\hbar\langle d_{\rm uc} \rangle D/a^2$. Since we do not
require static disorder and diffusion within the unit cell, the interpretation
of $g_{\rm uc}$ as a conductance is purely formal.

With respect to $g_{\rm uc}$ and $N$, we distinguish two regimes: For $g_{\rm
uc} \gg N^2$, the classical dynamics becomes ergodic before the energy-time
uncertainty relation allows to resolve the inter-band spacing $1/\langle
d_{\rm uc} \rangle$. The sampling by the discrete Bloch phases is then too
coarse to reveal the continuous bands underlying the discrete levels, and the
full spectrum appears as a superposition of $N$ independent spectra. The
condition $g_{\rm uc} \gg N^2$ can also be expressed as $2\pi/N \gg
\theta_{\rm corr}$, where $2\pi/N$ is the Bloch-phase spacing, and
$\theta_{\rm corr} = 2\pi\sqrt{\pi/g_{\rm uc}}$ is the spectral correlation
length with respect to variation of the Bloch phase \cite{tan}. Only for
$g_{\rm uc} \ll N^2$, the arrangement of levels in bands is felt by the
two-point correlations. A number of $N\theta_{\rm corr}/2\pi$ levels then
contribute coherently to the spectral correlations on time scales $\tau \leq
1$. In this case the second option in Eq.~(\ref{eq4}), the ergodic regime of
the classical dynamics, is irrelevant. In the space-time domain, this amounts
to the diffusion cloud still being well localized within the chain at $\tau =
1$. Increasing the chain length beyond $N = \sqrt{g_{\rm uc}}$ merely results
in a finer resolution of the bands.

\section{Quantum Ballistic Regime}
Equation~(\ref{eq4}) was derived using the diagonal approximation with respect
to the classical phases.  The periodic orbits occurring in the underlying
trace formula are those of the symmetry-reduced space. The break time beyond
which the diagonal approximation ceases to be valid is therefore the
Heisenberg time for the unit cell $t_{\rm H}$, or equivalently, $\tau = 1$.
This means that Eq.~(\ref{eq4}) describes only the spectral correlations on
scales of a typical inter-band spacing or larger. Therefore, we adopt a
different approach to $\widetilde K_n(\tau)$ for $\tau > 1$, corresponding to
energy scales of the inter-band spacing and below. Starting anew from the
definition (\ref{eq1}), we use Poisson resummation to replace the sum over $m$
by an integral over $\theta$,
\begin{equation}
\tilde a_n(\tau) = \frac{1}{2\pi} \sum_{\mu=-\infty}^{\infty}
\int_0^{2\pi} {\rm d}\theta\,
{\rm e}^{{\rm i}(n - \mu N)\theta}
\sum_{\alpha} {\rm e}^{-2\pi{\rm i}r_\alpha(\theta)\tau},
\label{eq5}
\end{equation}
where $r_{\alpha}(\theta) = \langle d_{\rm uc} \rangle E_{\alpha}(\theta)$.
For large $\tau$ the phase of the integrand is rapidly oscillating, and the
integration can be performed within an asymptotic approximation. Provided the
bandwidth is of the order of the inter-band spacing, this approach is
justified for times
% $\tau\appgtr 1$.
% $\tau\,\lower0.5ex\mbox{$\sim$}\kern-0.79em\raise0.5ex\mbox{$>$}\, 1$.
$\tau\stackrel{>}{\sim}1$. It will in fact be seen in Section
\ref{sec:large_N} that this approximation exactly reproduces the large $\tau$
behaviour in the limit of $N\rightarrow \infty$.  For simplicity we disregard
special cases like inflection points or higher-order extrema which can be
treated, e.g., by a uniform Bessel-function approximation
\cite{sti}. Saddle-point integration leads to the condition
\begin{equation}
2\pi r'_{\alpha}\left(\theta_j(\nu)\right) = \nu
\label{eq6}
\end{equation}
for points $\theta_j(\nu)$ of stationary phase, with $\nu = (n - \mu N)/\tau$.
Replacing dimensionless quantities by unscaled ones gives the equivalent
condition $v_{\alpha}(k) = (n - \mu N)a/t$, where $v_{\alpha}(k)$ is the group
velocity for the band ${\alpha}$ at quasimomentum $k = \hbar\theta/a$. It
expresses the ballistic motion of Bloch waves. For $\tau > 1$, the phases
$2\pi{\rm i}r_{\alpha}(\theta_j)\tau$ left by the saddle-point integration can
be considered random. Upon squaring to obtain the form factors, we therefore
drop the off-diagonal contributions and get
\begin{equation}
\widetilde K_n(\tau) = \frac{\gamma_n}{\tau}
\sum_{\mu=-\infty}^{\infty} F\left(\frac{n - \mu N}{\tau}\right),
\label{eq7}
\end{equation}
where $F(\nu) = (4\pi^2 \Delta r)^{-1}\sum_{\alpha} \sum_j |r''_{\alpha}
(\theta_j(\nu))|^{-1}$ is a positive function depending on $n$ and $\tau$ only
through $(n - \mu N)/\tau$. Similar to what we found for $\tau < 1$, the
$\widetilde K_n(\tau)$ can be interpreted as sums of winding-number
distributions $F(\nu(n,\tau))$, each of which is stretching from a center at
$n = \mu N$, but now as a {\it linear\/} function of time, so that the
variance $\sigma^2$ grows quadratically. They are normalized by the common
prefactor $1/\tau$. In order to determine the as yet unknown function $F(\nu)$,
we use a heuristic argument in the spirit of Refs.~\cite{dit1,arg,ber}, and
extrapolate both the semiclassical expression (\ref{eq4}) and Eq.~(\ref{eq7})
towards $\tau = 1$. Expanding
\begin{eqnarray}
\widetilde K_n^{({\rm sc})}(\tau) &=&
\frac{\gamma_n\tau}{N}\, {\cal G}^{({\rm mod}\, 1)}\left(\frac{n}{N},
\frac{g_{\rm uc}\tau}{\pi N^2}\right) \nonumber\\
&=& \gamma_n\sqrt{\frac{\tau}{2 g_{\rm uc}}} \sum_{\mu=-\infty}^{\infty}
\exp\left(-\frac{\pi(n-\mu N)^2}{2\tau g_{\rm uc}}\right),
\label{eq8}
\end{eqnarray}
we match Eq.~(\ref{eq4}) with Eq.~(\ref{eq7}) at $\tau = 1$,
\begin{equation}
\sum_{\mu=-\infty}^{\infty} F(n - \mu N) =
\frac{1}{\sqrt{2 g_{\rm uc}}}\sum_{\mu=-\infty}^{\infty}
\exp\left(-\frac{\pi(n-\mu N)^2}{2 g_{\rm uc}}\right).
\end{equation}
Comparing both sums term by term, we obtain $F(\nu) = (2g_{\rm
uc})^{-1/2}\exp(-\pi\nu^2/2g_{\rm uc})$ and thus, for the regime $\tau > 1$,
\begin{equation}
\widetilde K_n(\tau) = \frac{\gamma_n}{N}\, {\cal G}^{({\rm mod}\,
1)}\left(\frac{n}{N}, \frac{g_{\rm uc}\tau^2}{\pi N^2}\right).
\label{eq9}
\end{equation}
In the presence of additional symmetries, the spectral statistics can
be that of the GOE throughout the Brillouin zone, possibly with weak
localization enhancements in the vicinity of the symmetry points (cf.\
the discussion of the quantum kicked rotor below). In this case, an
analogous matching procedure applies. % We omit it for brevity.

For the long-time behaviour of the entire set of $\widetilde K_n(\tau)$,
Eq.~(\ref{eq9}) implies the following scenario: Initially,
$\widetilde{K}_0(\tau)$ decays as $1/\tau$ as long as only the term with
$\mu=0$ contributes significantly.  As terms with larger $\mu$ attain a
comparable magnitude, all $\widetilde K_n(\tau)$ approach an asymptotic
constant $\gamma_n/N$ (an exact evaluation \cite{dit2} gives a correction
$-2/N^2$ in the presence of the band symmetry $d_{-m}(E) = d_m(E)$, if $N$ is
even). The asymptotic domain is reached at $\tau \approx N \sqrt{\pi/g_{\rm
uc}}$. This is the effective Thouless time for ballistic spreading. It
corresponds to the time when the uncertainty relation allows the resolution of
the typical separation $2\pi/(N \theta_{\rm corr}\langle d_{\rm uc}\rangle)$
of neighbouring discrete levels. It will be shown in the sequel that the
expression (\ref{eq9}) reproduces the numerical data surprisingly well for
$\tau > 1$.
%%%%%%%%%%%%%%%%%%%%%%%%%%%%%%%%%%%%%%%%%%%%%%%%%%%%%%%%%%%%%%%%%%%%%
\begin{figure}
\centerline{\psfig{figure=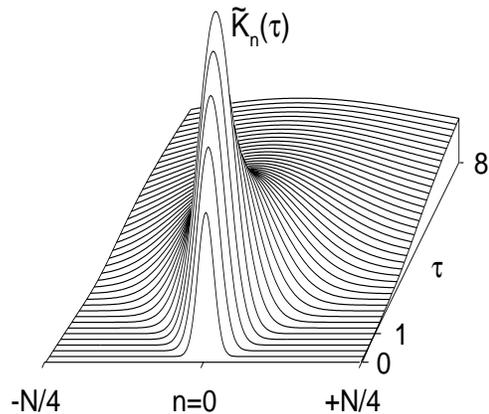,height=60mm,width=60mm,angle=-90}}
\begin{minipage}{8.0cm}\noindent
\caption{\label{fig-theo} 
Space-time dependence of the theoretical prediction for $\widetilde
K_n(\tau)$ according to Eqs.~(\protect\ref{eq4}) and (\protect\ref{eq9})
for $N=512$.}
\end{minipage}
\end{figure}
%%%%%%%%%%%%%%%%%%%%%%%%%%%%%%%%%%%%%%%%%%%%%%%%%%%%%%%%%%%%%%%%%%%%%
Fig.~\ref{fig-theo} summarizes our results for the time dependence of the
$\widetilde K_n(\tau)$. We can identify three distinct time regimes: a
semiclassical phase where the form factors reflect the classical diffusive
spreading, a regime dominated by quantal ballistic motion, and an asymptotic
regime reflecting the discreteness of the spectrum on the finest energy
scales. These long-time asymptotes vanish for large $N$ as $1/N$. At the same
time, the maximum of $\widetilde K_0(\tau)$ near $\tau=1$, all other
parameters being kept fixed, is proportional to $\theta_{\rm corr}$ and
independent of $N$. Therefore, in the regime $g_{\rm uc} \ll N^2$ where proper
bands exist, there is a crossover during which $\widetilde K_0(\tau)$ decays
by a factor of the order of $N\theta_{\rm corr}$. Thus, together with its rise
in the initial, semiclassical regime, $\widetilde K_0(\tau)$ attains a peak in
the vicinity of $\tau = 1$ which expresses the {\em clustering} of levels into
bands \cite{gei}. The stationary-phase condition (\ref{eq6}), with $\mu = 0$
(only this term contributes near $\tau = 1$), shows that the peak is
associated with the extremal points in the bands, that is, with the van-Hove
singularities \cite{ash} in the full spectral density $d(E)$.

Finally we note that for small $N$, in particular for $N = 2$, the
coarse-grained density relaxation does not follow a diffusion law if there is
no static disorder within the cells. An evaluation of the semiclassical and
the quantum domains along similar lines as sketched here, gives access to the
statistics of tunnel splittings in double billiards.

\section{The Large-$N$ Limit}
\label{sec:large_N}
In the present section we discuss the limit of $N\rightarrow\infty$. In this
limit, our results simplify considerably, since only the term $\mu=0$
contributes in Eqs.~(\ref{eq7}) and (\ref{eq8}). For short times
$\tau\stackrel{<}{\sim}1$ the $\widetilde{K}_n(\tau)$ can be replaced by a
single, diffusively spreading Gaussian as given in the second line of
Eq.~(\ref{eq4}). In the long-time regime $\tau \stackrel{>}{\sim} 1$, we
obtain
\begin{equation}
\widetilde{K}_n(\tau) = \frac{\gamma_n}{\sqrt{2 g_{\rm uc}} \tau}\,
\exp\left(-\frac{\pi}{2g_{\rm uc}}\left[{n\over\tau}\right]^2\right)\,.
\label{eq:limit}
\end{equation}
This implies in particular that $\widetilde K_0(\tau) = 1/\sqrt{2 g_{\rm
uc}}\tau$.

The same limiting behaviour has been calculated within the framework of a
nonlinear sigma model \cite{sim}. We shall now compare Eq.~(\ref{eq:limit}) to
the corresponding results in Ref.~\cite{sim}. We make use of the fact that the
Bloch phase $\theta$ can be viewed as a generalized Aharonov-Bohm flux and in
the present limit, becomes a continuous variable. Accordingly, we replace
$d_m(r)$ by $d(r,\theta)$. For simplicity, we restrict ourselves to the case
of broken time-reversal invariance throughout the Brillouin zone. This
corresponds to $\gamma_n = 1$ for all $n$. In order to enable a full
quantitative comparison with Ref.~\cite{sim}, we had to calibrate the
dimensionless conductance on basis of a common definition: In the present
limit, and for $\tau \gg 1$, we find the following relation between the
variance of the level velocities, the second moment of the $\widetilde
K_n(\tau)$ with respect to $n$, and the conductance,
\begin{equation}
4\pi^2\left\langle\left(\frac{{\rm d}r_{\alpha}(\theta)}
{{\rm d}\theta}\right)^2\right\rangle_{\alpha,\theta} =
\sum_{n=-\infty}^{\infty}\frac{n^2}{\tau^2}\widetilde K_n(\tau) =
\frac{g_{\rm uc}}{\pi}.
\end{equation}

In Ref.~\cite{sim}, the following expression is obtained for the flux-averaged
correlation function
\begin{eqnarray}
\lefteqn{\nonumber
\langle d(\overline r,\overline\theta) d(\overline r+r,\overline\theta+\theta) 
\rangle_{\overline r,\overline\theta} -1=}&&\\
\label{eq:altsh}
&&\frac{2 \pi}{g_{\rm uc}\theta^2}
\int_0^\infty \! \frac{{\rm d}\lambda}{\lambda}
\left\{
\exp\left(\displaystyle -{g_{\rm uc}\over \pi}
\left[{\theta\over 2\pi}\right]^2\left|\pi\lambda -\frac{\lambda^2}{2}
\right|\right)\right.\nonumber\\
&&-\left.\exp\left(\displaystyle -{g_{\rm uc}\over \pi}
\left[{\theta\over 2\pi}\right]^2\left|\pi\lambda +\frac{\lambda^2}{2}
\right|\right)\right\}\,\cos(r\lambda)
\end{eqnarray}
% \begin{eqnarray}
% \lefteqn{\nonumber
% \langle d(\overline r,\overline\theta) d(\overline r+r,\overline\theta+\theta) 
% \rangle_{\overline r,\overline\theta} -1}&&\\
% &=& 
% \label{eq:altsh}
% \frac{2 g_{\rm uc}}{\pi\theta^2}
% \int_0^\infty \! \frac{{\rm d}\lambda}{\lambda}\,\cos(r\lambda)
% %\\&\times &
% \left [
% \exp\left(\displaystyle -4\pi g_{\rm uc}\left|\pi\lambda -\frac{\lambda^2}{2}
% \right|\theta^2\right)\right.\nonumber\\
% &&-\left.\exp\left(\displaystyle -4\pi g_{\rm uc}\left|\pi\lambda +\frac{\lambda^2}{2}
% \right|\theta^2\right)
% \right ]
% \end{eqnarray}
An expression for $\widetilde K_n(\tau)$ is reached by Fourier transformation
with respect to $r$ and with respect to $\theta$ over the Brillouin zone.  The
discrete spatial Fourier transform cannot be applied directly to
Eq.~(\ref{eq:altsh}) since it violates periodicity in $\theta$. Instead,
exploiting the fact that this expression decays to zero for $\theta \to
\infty$, we approximate the Fourier sum over the Brillouin zone by a
continuous Fourier transformation over the whole real axis. We obtain
\begin{equation}
\widetilde K_n(\tau) = \frac{1}{2g_{\rm uc}\tau}
\left\{ \Phi_n(x_+)-\Phi_n(x_-)\right\}\,
\label{eq:altshfourier}
\end{equation}
with $\Phi_n(x) = \pi n\, \mbox{erf}(n/2\sqrt{x}) + 2\sqrt{\pi
x}\exp(-n^2/4x)$ and $x_\pm = (2\pi)^{-1} g_{\rm uc} |\tau\pm \tau^2|$.

\begin{figure}
\centerline{\psfig{figure=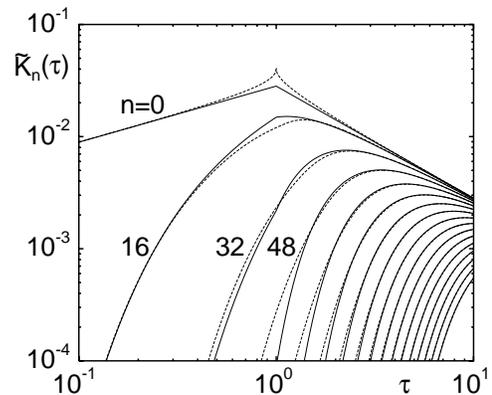,width=65mm}}
\vspace*{5mm}
\begin{minipage}{8.0cm}\noindent
\caption{\label{fig-altsh} Comparison of the form factors $\widetilde
K_n(\tau)$ in the limit $N \to \infty$, according to Eqs.~(\protect\ref{eq4})
and (\protect\ref{eq:limit}) (solid lines), to the corresponding functions as
implied by the results of Ref.~\protect\cite{sim}, cf.\
Eq.~(\protect\ref{eq:altsh}) (dashed). The graphs shown are for (top to
bottom) $n = 0$, 16, 32, $\ldots$, 256. The dimensionless conductance is
$g_{\rm uc} = 200\pi$, the same value as that underlying the data in
Fig.~\protect\ref{fig-num}a.
}
\end{minipage}
\end{figure}

We have checked analytically and numerically (Fig.~\ref{fig-altsh}) that both
for $\tau\to 0$ and $\tau\to\infty$, Eq.~(\ref{eq:altshfourier}) coincides
asymptotically with our results. Only in the vicinity of $\tau = 1$, there
exist deviations (e.g., from Eq.~(\ref{eq:limit}) we find $\widetilde K_0(1) =
1/\sqrt{2 g_{\rm uc}}$, while Eq.~(\ref{eq:altshfourier}) gives $\widetilde
K_0(1) = 1/\sqrt{g_{\rm uc}}$, cf.\ Fig.~\ref{fig-altsh}). The close agreement
between the respective results provides further support for the matching
procedure we used to connect the short- and long-time regimes. This agreement
is not surprising, since it has recently been shown \cite{muz,oded,jon} that
the random-matrix approach underlying Eq.~(\ref{eq:altsh}) also applies to
systems where the disorder is of dynamical origin.

%%%%%%%%%%%%%%%%%%%%%%%%%%%%%%%%%%%%%%%%%%%%%%%%%%%%%%%%%%%%%%%%%%%%%%
%%%%%%%%%%%%%%%%%%%%%%%%%%%%%%%%%%%%%%%%%%%%%%%%%%%%%%%%%%%%%%%%%%%%%%

\section{Numerical Results and Discussion}
\begin{figure}
\centerline{
\hspace*{10mm}
\psfig{figure=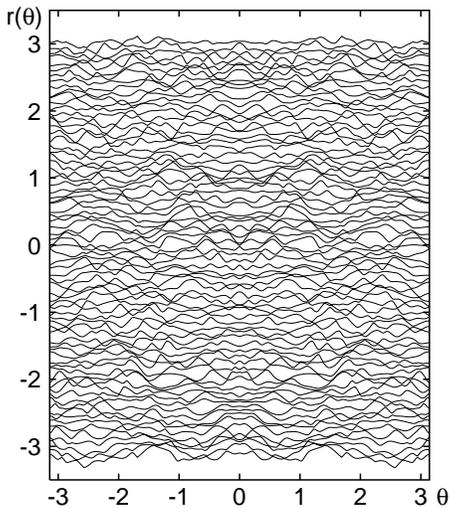,width=60mm}
\hspace*{4mm}
\vspace*{5mm}
}
\begin{minipage}{8.0cm}\noindent
\caption{\label{fig-spag} 
Quasienergy bands pertaining to chaotic Bloch states of the quantum kicked
rotor, for $k = 300$ and $\tau = 4\pi/75$, corresponding to a periodic
potential with a unit cell that accomodates 75 Bloch states.
}
\end{minipage}
\end{figure}

In the remainder, we compare the theory sketched in the previous paragraphs to
two prototypical yet quite different models. Our first example is the kicked
rotor on a torus \cite{izr}, defined by its Hamiltonian
\begin{equation}
H(l,\vartheta;t) ={(l-\lambda)^2\over 2} + V_{\alpha,k}(\vartheta)
\sum_{m=-\infty}^{\infty} \delta(t-m\tau)\,.
\end{equation}
It is periodically time dependent, so that spectrum and eigenstates are
adequately discussed in terms of quasienergies and Floquet states,
respectively. The kicked rotor attains periodicity also in the
angular-momentum variable $l$ if the parameter $\tau$, an effective quantum of
action, is chosen as $\tau = 4\pi p/q$, with $p$, $q$ coprime (here we
restrict them further to $p = 1$ and $q$ odd). The unit cell then accomodates
$q$ quanta of angular momentum. The number of quasienergy bands (for a typical
sample, see Fig.~\ref{fig-spag}) is also $q$. It is analogous to the total
number of bands $\Delta r$ introduced above.

The quasienergies are obtained by diagonalizing the symmetry-projected Floquet
operator at Bloch phase $\theta_m$ \cite{izr},
% \begin{eqnarray}
% \langle\,l'\,|\widehat U_m|\,l\,\rangle &=&
% \exp\left(-2\pi{\rm i}\frac{p}{q} (l-\lambda)^2\right)\nonumber\\
% &\times& \frac{1}{q} \sum_{n=0}^{q-1}
% \exp\left(-{\rm i}V_{\alpha,k}\left(\frac{\theta_m+2\pi n}{q}\right)\right)
% \exp\left({\rm i}(l-l')\frac{\theta_m+2\pi n}{q}\right).
% \label{eq10}
% \end{eqnarray}
\begin{eqnarray}
&& \langle\,l'\,|\widehat U_m|\,l\,\rangle =
\exp\left(-2\pi{\rm i}\frac{p}{q} (l-\lambda)^2\right) \nonumber\\
&&\quad \times \frac{1}{q} \sum_{n=0}^{q-1}
{\rm e}^{-{\rm i}V_{\alpha,k}([\theta_m+2\pi n]/q)}
{\rm e}^{{\rm i}(l-l')(\theta_m+2\pi n)/q}.
\label{eq10}
\end{eqnarray}
For the kicked rotor on a torus, the number of unit cells is simply determined
by the number $N$ of Bloch phases where Eq.~(\ref{eq10}) is evaluated. This
additional parameter is independent of $q$, the number of bands. The total
number of levels in the spectrum is therefore $Nq$.

%%%%%%%%%%%%%%%%%%%%%%%%%%%%%%%%%%%%%%%%%%%%%%%%%%%%%%%%%%%%%%%%%%%%%%
%%%%%%%%%%%%%%%%%%%%%%%%%%%%%%%%%%%%%%%%%%%%%%%%%%%%%%%%%%%%%%%%%%%%%%
Besides being periodic in $l$ and $l'$, the Floquet matrix (\ref{eq10}) allows
for several twofold symmetries \cite{izr,blu,cas}. In order to break them in a
controlled manner without significantly altering the classical diffusion, an
angular-momentum shift by $\lambda$ has been introduced, and the potential has
been chosen as \cite{blu}
\begin{equation}
V_{\alpha,k}(\vartheta) = k\left[\cos(\alpha\frac{\pi}{2})\cos\,\vartheta +
\frac{1}{2}\sin(\alpha\frac{\pi}{2})\sin\,2\vartheta\right].
\label{eq11}
\end{equation}

For $\alpha = \lambda = m = 0$, which classically corresponds to the common
standard map, the kicked rotor possesses two independent twofold symmetries, a
unitary one, the parity ${\sf P}$: $l \to -l$, $\vartheta \to 2\pi -
\vartheta$, $t \to t$, and an antiunitary one, time reversal ${\sf T}$: $l \to
-l$, $\vartheta \to \vartheta$, $t \to -t$. ${\sf P}$ invariance is broken for
$m \neq 0$, $L/2$, i.e., off the band center and edges. Therefore, for $\alpha
= \lambda = 0$, the quasienergy statistics corresponds to the superposition of
two independent COE's at $m = 0$, $\pm L/2$, and to a single COE
elsewhere. Choosing $\alpha \neq 0$ breaks ${\sf P}$ also at the symmetry
points and at the same time, lifts the band symmetry so that COE statistics
becomes valid throughout the Brillouin zone. If $\lambda$ takes a non-integer
value, ${\sf T}$ invariance, which amounts to $\langle\, -l'\,|\widehat
U_m|\,-l\,\rangle = \langle\, l'|\widehat U_m|\,l\,\rangle$, is broken. In
this case, the quasienergy statistics is that of the CUE everywhere, or that
of two superposed CUE's at the symmetry points if $\alpha = 0$.

%%%%%%%%%%%%%%%%%%%%%%%%%%%%%%%%%%%%%%%%%%%%%%%%%%%%%%%%%%%%%%%%%%%%%%
\begin{figure}
\centerline{\psfig{figure=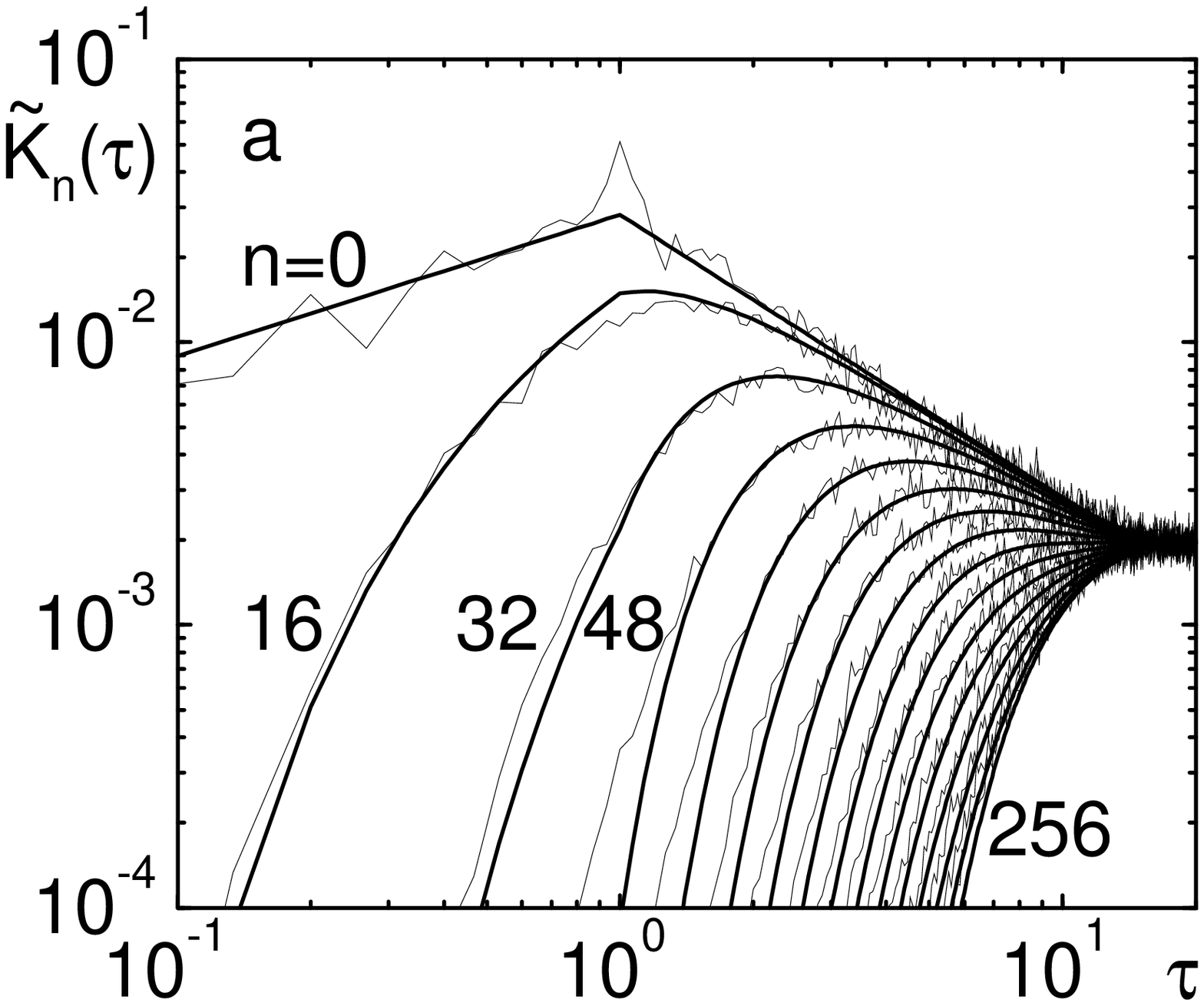,width=65mm}}
\vspace*{2mm}
\centerline{\psfig{figure=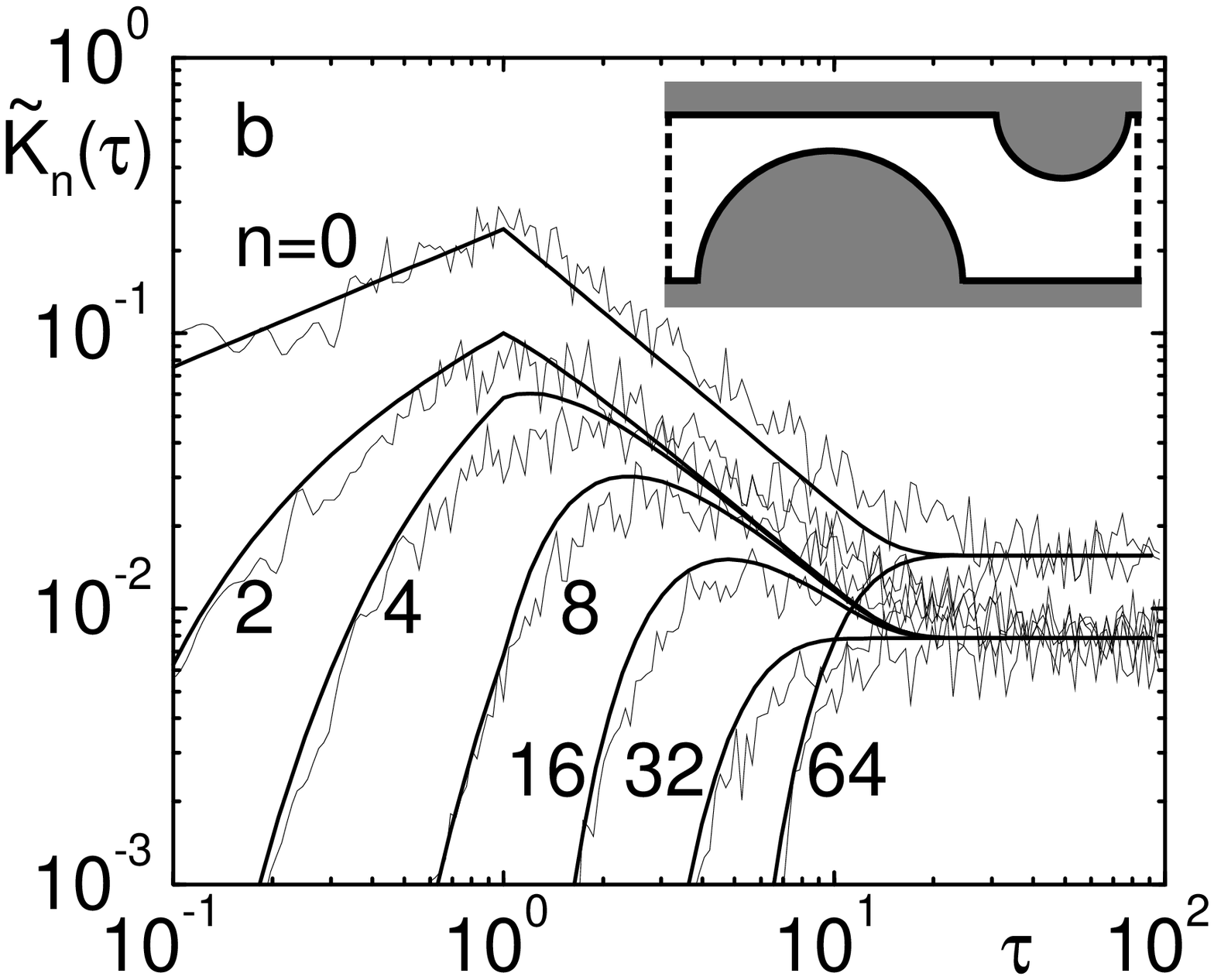,width=65mm}}
\vspace*{1mm}
\begin{minipage}{8.0cm}\noindent
\caption{\label{fig-num} (a) Form factors $\widetilde K_n(\tau)$ for the
kicked rotor on a torus at (top to bottom) $n = 0$, 16, 32, $\ldots$, 256 ($N
= 512$), compared to the theory (heavy lines). The parameters determining
$g_{\rm uc}$ are, in the notation of Ref.~\protect\cite{dit1}, $\tau
=4\pi/225$ (dimensionless quantum of action, not to be confounded with the
scaled time used in this letter) and $k = 300$, giving $g_{\rm uc} =
200\pi$. All twofold antiunitary symmetries are broken.
(b) is analogous to (a) but for a billiard chain composed of $N=128$ unit
cells (inset). The winding numbers shown are $n=0$, 2, 4, 8, 16, 32, 64. The
only free parameter of the theory, $g_{\rm uc}$, was determined by an
independent classical simulation~\protect\cite{dit2}.  }
\end{minipage}
\end{figure}
%%%%%%%%%%%%%%%%%%%%%%%%%%%%%%%%%%%%%%%%%%%%%%%%%%%%%%%%%%%%%%%%%%%%%%
Sufficiently large spectral data sets have been generated by varying $k$ over
small intervals so that the classical diffusion constant $D \approx k^2/2$ is
not changed appreciably. The parameters $N$, $q$, and $k$ were chosen such
that the scale of localization due to disorder within the unit cell
\cite{tan,izr} by far exceeds the cell size. In Fig.~\ref{fig-num}a, we show,
for both twofold symmetries simultaneously broken ($\alpha = 0.2$, $\lambda =
0.5$), the set of $\widetilde K_n(\tau)$ as a function of $\tau$ at selected
equidistant values of $n$, and compare with our theory, Eqs.~(\ref{eq4}) and
(\ref{eq9}). The three time regimes and the corresponding power-law time
dependences of $\widetilde K_0(\tau)$ can clearly be distinguished. Comparing
with the nonlinear-sigma-model results \cite{sim} (Eq.~(\ref{eq:altshfourier})
and the dashed lines in Fig.~\ref{fig-altsh}), we find that near $\tau = 1$,
they reproduce the data even better, including an unexpected feature like the
sharp peak at $n = 0$, $\tau = 1$. The saturation regime $\tau
\stackrel{>}{\sim} N\sqrt{\pi/g_{\rm uc}}$, however, is not contained in
Eq.~(\ref{eq:altshfourier}).

Our second example is a billiard chain composed of unit cells as shown in the
inset in Fig.~\ref{fig-num}b. The wave functions satisfy the Helmholtz equation
%\begin{equation}
%(\Delta+k^2)\Psi({\bf q})=0\,,
%\end{equation}
augmented with Dirichlet boundary conditions on the channel walls and periodic
boundary conditions along the channel after $N$ unit cells.  Provided that
trajectories traversing the unit cell without hitting the obstacles are
excluded by an appropriate geometry, the classical dynamics is diffusive on
scales larger than the size of the unit cell. The energy bands can be found
using the scattering approach to billiard quantization, see Ref.~\cite{dit2}
for details of the method. The shape of the unit cell is chosen such that
there are no other unitary symmetries besides the discrete translation
invariance. However, due to time-reversal invariance, the bands are symmetric
with respect to $\theta = 0$ and $\pi$, such that $\gamma_0 = \gamma_{N/2} =
2$ and else $\gamma_n = 1$. Fig.~\ref{fig-num}b, which is analogous to
Fig.~\ref{fig-num}a, clearly shows the corresponding enhancements of
$\widetilde K_0(\tau)$ and $\widetilde K_{N/2}(\tau)$ throughout the three
regimes. Correspondingly, the spectral statistics for fixed Bloch number
follows the COE at the center and the edge of the Brillouin zone and
approaches CUE in between \cite{dit2}.

Our theoretical results reproduce remarkably well the numerical data obtained
for the two models. This lends support to the approximations underlying the
theory and corroborates its conclusions. The large-$N$ limit is of particular
interest, because it provides a very sensitive tool for studying the
transition to Anderson localization when the periodicity is disrupted by
disorder. In the latter case, the level statistics approaches the Poissonian
limit, and the corresponding spectral form factor approaches the constant
value $1$ in the long-time limit. This is in a marked difference from the
periodic case, where $\widetilde K_0(\tau) \approx 1/\tau $ for $\tau >
1/N$. The transition between these two rather distinct asymptotics occurs as a
smooth function of the degree of disorder. In order to understand this
transition in semiclassical terms \cite{dit1,arg}, one may discuss the effect
of weak disorder through the small random differences between the phases
assigned to periodic orbits which were related by symmetry in the absence of
disorder. This approach is similar to the semiclassical discussion of the
response of the weak-localization peak to a weak magnetic field \cite{blu}. A
detailed analysis of this scenario will be left for further study.

\section*{Acknowledgements}
The research reported in this work was supported by grants from the Minerva
Center for Nonlinear Physics. Three of us (TD, BM, HS) would like to thank
the Weizmann Institute of Science, Rehovot, and US would like to thank the Max
Planck Institute for Physics of Complex Systems, Dresden, for the kind
hospitality enjoyed during several visits.

\newcommand{\acp}[1]{Adv.\ Chem.\ Phys.\ {\bf #1}}
\newcommand{\ama}[1]{Ann.\ Math.\ {\bf #1}}
\newcommand{\ana}[1]{Annals N.Y.\ Acad.\ Sci.\ {\bf #1}}
\newcommand{\apb}[1]{Act.\ Phys.\ Pol.\ B {\bf #1}}
\newcommand{\aph}[1]{Adv.\ Phys.\ {\bf #1}}
\newcommand{\apn}[1]{Ann.\ Phys.\ (N.Y.) {\bf #1}}
\newcommand{\cha}[1]{Chaos {\bf #1}}
\newcommand{\cmp}[1]{Commun.\ Math.\ Phys.\ {\bf #1}}
\newcommand{\csf}[1]{Chaos, Solitons \& Fractals {\bf #1}}
\newcommand{\dan}[1]{Dokl.\ Akad.\ Nauk SSSR {\bf #1}}
\newcommand{\epl}[1]{Europhys.\ Lett.\ {\bf #1}}
\newcommand{\hpa}[1]{Helv.\ Phys.\ Acta {\bf #1}}
\newcommand{\ibm}[1]{IBM J.\ Res.\ Dev.\ {\bf #1}}
\newcommand{\jch}[1]{J.\ Phys.\ Chem.\ {\bf #1}}
\renewcommand{\jcp}[1]{J.\ Chem.\ Phys.\ {\bf #1}}
\newcommand{\jel}[1]{JETP Lett.\ {\bf #1}}
\newcommand{\jmp}[1]{J.\ Math.\ Phys.\ {\bf #1}}
\newcommand{\jsp}[1]{J.\ Stat.\ Phys.\ {\bf #1}}
\newcommand{\jpa}[1]{J.\ Phys.\ A {\bf #1}}
\newcommand{\jpc}[1]{J.\ Phys.\ C {\bf #1}}
\newcommand{\ncl}[1]{Lett.\ Nuovo Cimento {\bf #1}}
\newcommand{\ndy}[1]{Nonlinear Dynamics {\bf #1}}
\newcommand{\nly}[1]{Nonlinearity {\bf #1}}
\newcommand{\pga}[1]{Physica {\bf #1A}}
\newcommand{\pgb}[1]{Physica {\bf #1B}}
\newcommand{\pgd}[1]{Physica {\bf #1D}}
\newcommand{\pha}[1]{Physica A {\bf #1}}
\newcommand{\phb}[1]{Physica B {\bf #1}}
\newcommand{\phd}[1]{Physica D {\bf #1}}
\newcommand{\phm}[1]{Philos.\ Mag.\ {\bf #1}}
\newcommand{\phr}[1]{Phys.\ Rep.\ {\bf #1}}
\newcommand{\plt}[1]{Phys.\ Lett.\ {\bf #1}}
\newcommand{\pla}[1]{Phys.\ Lett.\ A {\bf #1}}
\newcommand{\prv}[1]{Phys.\ Rev.\ {\bf #1}}
\renewcommand{\pra}[1]{Phys.\ Rev.\ A {\bf #1}}
\renewcommand{\prb}[1]{Phys.\ Rev.\ B {\bf #1}}
\renewcommand{\prd}[1]{Phys.\ Rev.\ D {\bf #1}}
\renewcommand{\pre}[1]{Phys.\ Rev.\ E {\bf #1}}
\renewcommand{\prl}[1]{Phys.\ Rev.\ Lett.\ {\bf #1}}
\newcommand{\psa}[1]{Proc.\ R.\ Soc.\ A {\bf #1}}
\newcommand{\pta}[1]{Phil.\ Trans.\ R.\ Soc.\ A {\bf #1}}
\newcommand{\pts}[1]{Prog.\ Theor.\ Phys.\ Suppl.\ {\bf #1}}
\renewcommand{\rmp}[1]{Rev.\ Mod.\ Phys.\ {\bf #1}}
\newcommand{\rpp}[1]{Rep.\ Prog.\ Phys.\ {\bf #1}}
\newcommand{\sci}[1]{Science {\bf #1}}
\newcommand{\spd}[1]{Sov.\ Phys.\ Dokl.\ {\bf #1}}
\newcommand{\spj}[1]{Sov.\ Phys.\ JETP {\bf #1}}
\newcommand{\ssc}[1]{Sol.\ Stat.\ Comm.\ {\bf #1}}
\newcommand{\ssr}[1]{Sov.\ Sci.\ Rev.\ {\bf #1}}
\newcommand{\tam}[1]{Trans.\ Am.\ Math.\ Soc.\ {\bf #1}}
\newcommand{\tmf}[1]{Teor.\ Mat.\ Fiz.\ {\bf #1}}
\newcommand{\tmp}[1]{Theor.\ Math.\ Phys.\ {\bf #1}}
\newcommand{\tms}[1]{Trans.\ Moscow Math.\ Soc.\ {\bf #1}}
\newcommand{\vdp}[1]{Verh.\ Dt.\ Phys.\ Ges.\ {\bf #1}}
\newcommand{\zet}[1]{Zh.\ Eksp.\ Teor.\ Fiz.\ {\bf #1}}
\newcommand{\znf}[1]{Z.\ Naturforsch.\ {\bf #1}}
\newcommand{\zpb}[1]{Z.\ Phys.\ B {\bf #1}}
\newcommand{\zpd}[1]{Z.\ Phys.\ D {\bf #1}}

\end{multicols}

\end{document}